\documentclass[%
aip,
jmp,%
amsmath,amssymb,
reprint,%
nofootinbib
]{revtex4-2}

\usepackage{graphicx}
\usepackage{dcolumn}
\usepackage{bm}

\usepackage[utf8]{inputenc}
\usepackage[T1]{fontenc}
\usepackage{mathptmx}
\usepackage{etoolbox}
\usepackage[usenames,dvipsnames]{xcolor}

\makeatletter
\def\@email#1#2{%
 \endgroup
 \patchcmd{\titleblock@produce}
  {\frontmatter@RRAPformat}
  {\frontmatter@RRAPformat{\produce@RRAP{*#1\href{mailto:#2}{#2}}}\frontmatter@RRAPformat}
  {}{}
}%
\makeatother

\begin{document}


\title{Optoelectrical nanomechanical resonators made from multilayered 2D materials}

\author{Joshoua Condicion Esmenda}
\thanks{Equally contributing authors.}
 \email{jesmenda@gate.sinica.edu.tw.}
 \affiliation{Department of Engineering and System Science, National Tsing Hua University, Hsinchu 30013, Taiwan}
 \affiliation{Nano-Science and Technology Program, Taiwan International Graduate Program, Academia Sinica, Taipei 11529, Taiwan}
 \affiliation{Institute of Physics, Academia Sinica, Taipei 11529, Taiwan}
 
\author{Myrron Albert Callera Aguila}
\thanks{Equally contributing authors.}
 \affiliation{Department of Engineering and System Science, National Tsing Hua University, Hsinchu 30013, Taiwan}
 \affiliation{Nano-Science and Technology Program, Taiwan International Graduate Program, Academia Sinica, Taipei 11529, Taiwan}
 \affiliation{Institute of Physics, Academia Sinica, Taipei 11529, Taiwan}
 \email{maguila@gate.sinica.edu.tw}
 
\author{Jyh-Yang Wang}
 \affiliation{Institute of Physics, Academia Sinica, Taipei 11529, Taiwan}
 
\author{Teik-Hui Lee}
 \affiliation{Institute of Physics, Academia Sinica, Taipei 11529, Taiwan}
 
\author{Yen-Chun Chen}
 \affiliation{Institute of Physics, Academia Sinica, Taipei 11529, Taiwan}
 
\author{Chi-Yuan Yang}
 \affiliation{Department of Physics, National Taiwan University, Taipei 10617, Taiwan}
 \affiliation{Institute of Physics, Academia Sinica, Taipei 11529, Taiwan}

\author{Kung-Hsuan Lin}
 \affiliation{Institute of Physics, Academia Sinica, Taipei 11529, Taiwan}
 
\author{Kuei-Shu Chang-Liao}
 \affiliation{Department of Engineering and System Science, National Tsing Hua University, Hsinchu 30013, Taiwan}

\author{Sergey Kafanov}
\affiliation{Department of Physics, Lancaster University, LA1 4YB, Lancaster, United Kingdom}

\author{Yuri Pashkin}
\affiliation{Department of Physics, Lancaster University, LA1 4YB, Lancaster, United Kingdom}

\author{Chii-Dong Chen}
 \affiliation{Institute of Physics, Academia Sinica, Taipei 11529, Taiwan}
  \email{chiidong@phys.sinica.edu.tw}

\begin{abstract}
Studies involving nanomechanical motion have evolved from its detection and understanding of its fundamental aspects to its promising practical utility as an integral component of hybrid systems.  Nanomechanical resonators' indispensable role as transducers between optical and microwave fields in hybrid systems, such as quantum communications interface, have elevated their importance in recent years.  It is therefore crucial to determine which among the family of nanomechanical resonators is more suitable for this role.  Most of the studies revolve around nanomechanical resonators of ultrathin structures because of their inherently large mechanical amplitude due to their very low mass.  Here, we argue that the underutilized nanomechanical resonators made from multilayered two-dimensional (2D) materials are the better fit for this role because of their comparable electrostatic tunability and larger optomechanical responsivity.  To show this, we first demonstrate the electrostatic tunability of mechanical modes of a multilayered nanomechanical resonator made from graphite.  We also show that the optomechanical responsivity of multilayered devices will always be superior as compared to the few-layer devices.  Finally, by using the multilayered model and comparing this device with the reported ones, we find that the electrostatic tunability of devices of intermediate thickness is not significantly lower than that of ultrathin ones.  Together with the practicality in terms of fabrication ease and design predictability, we contend that multilayered 2D nanomechanical resonators are the optimal choice for the electromagnetic interface in integrated quantum systems.
\end{abstract}

\keywords{Plate nanomechanical resonators, Electrostatic tunability, Tensioned thin plate, Nanoelectromechanical systems (NEMS).}

\maketitle

\section*{Introduction}
Interaction between mechanical motion and electromagnetic radiation has been greatly studied in recent years because of its many applications in laser science\cite{dennis2015compact,van2012ultracompact,pruessner2016broadband}, sensing\cite{mamin2001sub,lahaye2004approaching,yang2006zeptogram}, and quantum information processing\cite{hease2020bidirectional,forsch2020microwave}. Mechanical resonators are the best candidates for playing the role of converters of electromagnetic waves of one wavelength to another\cite{midolo2018nano}.  For example, in quantum communication, mechanical resonators in the nanoscale act as intermediaries between optical and microwave photons\cite{aspelmeyer2014}. For large-scale quantum networks, linking quantum processors with low-loss optical fibers can be done by using mechanical resonators as microwave-to-optical converters, that exhibit bidirectional, coherent, and efficient transduction\cite{Andrews2014}. Furthermore, introducing a level of control by tuning the mechanical resonance can be used to enhance the coupling to either degree of freedom. It is therefore important that these mechanical resonators have the versatility in both design and mechanical tunability. Nanomechanical resonators (NMRs) made from two-dimensional (2D) materials offer flexibility, not only because of the variety of materials to choose from, but,  more importantly,  due to their high elasticity and low mass\cite{Bunch2007,luo2018strong,zhang2020coherent}.
\\\indent
Much of the study involving 2D NMRs gear towards ultrathin structures approaching the atomic scale\cite{Bunch2007,Bagci2014,DeAlba2016} because ultrathin structures have low mass that enables large displacement, which is integral to the mechanical resonators’ coupling mechanism\cite{palomaki2013coherent,aspelmeyer2014}. However, as transducers for optical fields, structures made from multiple layers become more appealing as thicker plates provide higher light refraction\cite{li2014measurement}. Furthermore, it is known that ultrathin resonators are more susceptible to built-in strain as compared to multilayer ones, that ultimately design predictability. Studies of multilayered 2D NMRs\cite{ben2021magnetic} interacting with light are few\cite{lee2013high,wang2015black,zheng2017hexagonal,islam2018anisotropic,wang2019hexagonal,ma2014fiber} and those with tuning by electrical means are even fewer\cite{wang2016resolving}.
\\\indent
In this work, we present an NMR made from multiple layers of graphene in a Fabry-Perot structure characterized through laser interferometry. Its tunability by electrostatic means for the mechanical modes is demonstrated.  The enhancement of the optomechanical responsivity as it relates to the number of layers is also shown. To better understand the effect of geometry and material properties, a model for multilayered resonators is constructed.  Finally,   the resonance frequency tunability of the device is compared to that of ultrathin resonators of a similar structure.  By looking at the electrostatic tunability of multilayered 2D NMRs, we can see that they are the optimal choice for transducers,  especially between optical and electrical fields.  Their significant role in quantum communication,  for example,  will make this choice of device very critical.
\section*{Results}
\subsection*{\label{sec:level2}Description and characterization of the device}
\indent
Figure 1a shows an optical image of the sample containing a multilayered graphene drum,  which we will call Device A throughout this paper,  and it was determined to have 50 layers\cite{Myrron2020}.  Figure 1b shows a schematic cross-section of the chip along the white dashed line shown in Figure 1a,  where the flake is placed on top of the spacer.  Figure 1c shows a schematic of how the electrostatic force $F_{dc}$ applied to the drum is counterbalanced by the total radial tension $T$.  The motion of the mechanical drum is then detected using the laser interferometry technique as illustrated in Figure 1d (details of the fabrication, mechanical detection, and actuation are described in the Supplementary Information).  Peaks were detected at 5.7 MHz,  10.9 MHz,  and 12.0 MHz for the fundamental mode $f_0$,  and the higher modes,  $f_1$,  and $f_2$.  Next we study how the peak position and amplitude depend on the applied DC voltage.
\subsection*{\label{sec:level2}Electrostatic Tunability}
\indent
Figure 2 shows the measured response of the device for its first three mechanical modes at different values of the DC voltage.  Each response curve exhibits the Lorentzian shape typical for a damped driven harmonic oscillator.  The response magnitude is converted to the mechanical amplitude by determining the device's optomechanical responsivity,  the details of which will be discussed in the latter part of this work. The response curves show that the mechanical amplitude increases as the absolute value of the DC voltage increases.  The back plane shows this increasing trend more clearly with the maximum amplitude values of about 880 pm,  14 pm,  and 5 pm for $f_0$,  $f_1$ and $f_2$,  respectively,  at the maximum applied negative DC voltage.  We also see the resonance peaks shifting to higher frequencies as the DC voltage increases. The bottom plane provides a guide to see this trend in peak frequency shift,  $\Delta f$,  with respect to the resonance frequency at $V_{dc}$ $\approx$ 0, with all of the modes having $\Delta f$ of about 0.2 MHz. 
\\\indent
To analyze these increasing trends in both mechanical amplitude and frequency in detail,  we plot the peak amplitude and the resonance frequency as a function of the DC voltage,  as shown in Figure 3.  In Figure 3a,  we see that all mechanical modes show an increasing frequency shift that follows the shape of a semi-circle with the form $f = \sqrt{c_0+c_1(V_{dc}-c_2)^2}$, where $c_0$,  $c_1$,  and $c_2$ are positive constants used as fitting parameters.  In fact,  we will use a model with the same form to fit the trend and extract the device parameters from it.  Figure 3b,  on the other hand,  shows the increasing trend of the mechanical amplitude and they all exhibit a linear behavior.  As expected,  the mechanical amplitude of the higher modes are smaller than that of the fundamental mode.  However,  there is an observed asymmetry of the mechanical amplitude of all modes at negative and positive values of the DC voltage with respect to $V_{dc}=c_2$ as shown in Figure 3b.   Furthermore,  this asymmetry is more obvious in the higher modes $f_1$ and $f_2$ than in the fundamental mode $f_0$.  In order to offer an explanation as to why this occurs,  formulation of the mechanical frequency and amplitude dependence on the DC voltage is needed.  Before that,  however,  we must first be able to describe how the measured response magnitude of the device is converted to the actual mechanical amplitude,  as this demonstrates how well the device is able to calibrate the mechanical motion.
\section*{Discussion}
\subsection*{\label{sec:level2}Optomechanical Responsivity}
\indent
As explained earlier, the information about mechanical motion is carried by the modulated laser beam reflected from the device. The modulation amplitude can be converted to the amplitude of mechanical vibrations by obtaining the device's optomechanical responsivity.  In order to do this,  the total reflectance $R$ is differentiated with respect to the displacement from equilibrium $z$.  However,  calculation of the total reflectance is not trivial because each of the layers has its own refraction index that is layer thickness dependent.  For this work, however, we will only focus on the graphite flake device's optomechanical responsivity and the multilayer interface approach is explained in detail elsewhere\cite{Myrron2020}.  Figure 4 shows the result of the approach, where the optomechanical responsivity (OR),  $dR/dz$,  is calculated over a range of the vacuum gap $g_0$ and number of graphene layers.  This allows conversion of the measured response to mechanical amplitude and provides understanding of the effect of key device parameters such as thickness and vacuum gap on the overall optomechanical responsivity.  From Figure 4a,  we can say that there are certain combinations of the thickness and vacuum gap that give an optimal optomechanical responsivity.  In particular,  for the given vacuum gap of 285 nm, the highest responsivity is produced by the plate which is about 20 layers thick,  as illustrated in Figure 4b.  This peak of the optomechanical responsivity can be intuitively explained by two mechanisms of light and device interaction: reflection and absorption\cite{wang2016interferometric}. The increase of optomechanical responsivity from one layer to a few number of layers is due to the increased multiple reflection, and consequently, increased constructive interference. However, as the number of layers continues to increase, the effect of absorption becomes more significant and effectively decreases the overall total reflection and optomechanical responsivity. In fact, if Figure 4a is extended to a very large number of layers, the interference effects will be dominated by the total reflection from the drumhead only for all values of the vacuum gap. Moreover,  while the vacuum gap has a periodic influence on the optomechanical responsivity,  it is clear that the multiple layers of the drumhead material will have better optomechanical responsivity than that of the few layers.  Now that we understand the importance of the number of layers for the optomechanical responsivity,  let us see how the thickness influences frequency tunability of the device by looking at its fitting model.
\subsection*{\label{sec:level2}Multilayered NMR model}
\indent
For multilayer structures,  it is common to discuss whether the resonator is in the modulus dominated regime,  commonly referred to as a plate or disk,  or in the tension dominated regime,  when the resonator is called a membrane.  The transition between these two regimes,  however,  can be described by the vibrational motion of a plate under tension in the absence of an external force with the following equation\cite{wah1962vibration}:
\begin{equation}
\nabla ^4(z) -  \frac{T}{D} \nabla ^2(z)+ \frac{\rho}{D}  \frac{\partial^2z}{\partial t^2} = 0.      
\label{eq:Equation_1}
\end{equation}
The symbols $z$,  $T$,  $D$ and $\rho$ are the mechanical displacement,  flexural rigidity,  tension and material density,  respectively.  A more accurate model of the rigidity would be to include the interlayer shear\cite{chen2015bending,liu2011interlayer} but,  for simplicity,  it will not be included at the moment to maintain focus on the effect of the induced strain from the electrostatic force.  From here, the resonance frequency for the $(m,n)$ mode,  where $m$ and $n$ represent the number of nodal diameters and nodal circles,  respectively,  and both rigidity and tension are taken into account,  is given by\cite{sazonova2006tunable}:
\begin{equation}
f_{mn} =  \sqrt{f_{\mathbf{p},mn}^2+f_{\mathbf{m},mn}^2}.
\label{eq:Equation_2}
\end{equation}
where $f_{\mathbf{p},mn}$ and $f_{\mathbf{m},mn}$ are the plate and membrane resonance frequencies,  respectively.  We can write down explicitly the corresponding expressions for the membrane and plate resonance frequencies as follows\cite{Silvan2001}:
\begin{equation}
f_{\mathbf{p},mn} = \sqrt{\frac{\lambda_{\mathbf{p},mn}^4D}{4\pi^2r^4\rho h}},
\label{eq:Equation_3}
\end{equation}
\begin{equation}
f_{\mathbf{m},mn} = \sqrt{\frac{\lambda_{\mathbf{m},mn}^2\sigma}{4\pi^2r^2\rho}},
\label{eq:Equation_4}
\end{equation}
where $\sigma$ is the tensile stress,  $h$ is the thickness of the drumhead, and $\lambda_p$ and $\lambda_m$ are the respective eigenmode numbers for the corresponding plate and membrane resonances. The tensile stress can be further described in terms of the total radial tension $T$ to counteract $F_{dc}$, where $F_{dc}$ is equal to  0.5$\frac{dC}{dz}(V_{dc}-V_0)^2$,  with $C$ representing the capacitance between the drumhead and bottom electrode,  $T_{initial}$ the initial tension,  $V_{0}$ the offset voltage due to the work function difference of the drumhead and contacts materials\cite{sazonova2006tunable},  and $\frac{dC}{dz}$ is the differential motional capacitance:
\begin{equation}
\sigma = \frac{T}{2\pi rh} =  \frac{T_{initial}}{2\pi rh} + \frac{dC}{dz} \frac{ (V_{dc}-V_{0})^2}{4\pi rh}.
\label{eq:Equation_5}
\end{equation}
The final equation for the frequency of a tensioned plate takes the form:
\begin{equation}
f_{mn} = \sqrt{\frac{\lambda_{\mathbf{p},mn}^4D}{4\pi^2r^4\rho h}+\frac{\lambda_{\mathbf{m},mn}^2T_{initial}}{8\pi^3r^3h\rho}+\frac{dC}{dz}\frac{\lambda_{\mathbf{m},mn}^2(V_{dc}-V_{0})^2}{16\pi^3r^3h\rho}}.
\label{eq:Equation_6}
\end{equation}
Now that we have a model that approximates the contributions of both rigidity and tension in Equation 6,  we can use it to fit the data of Figure 3a.  We then can retrieve three parameters: the offset voltage, the initial tension $T_{initial}$ and the differential motional capacitance.  From the fit,  the offset voltage is -103 mV while the initial tension is 47.38 mN.  From this value of initial tension,  we evaluate the parameter $r\sqrt{\frac{T}{D}\ }$ to be about 0.11.  This indicates that the resonator behaves more like a plate\cite{van2010modelling}.  The differential motional capacitance is typically hard to determine because of the inherent non-uniform deformation present in all resonators.  From the fit,  the differential motional capacitance is determined to be 417.3 nF/m.  Also,  the noticeable difference in curvature between the fitting curves for $f_0$, $f_1$ and $f_2$ can be attributed to the difference in their $\lambda$'s.  It is expected that the $\lambda$ for $f_0$ should be different from $f_1$ and $f_2$ since they represent $(0,1)$ and $(1,1)$ mode,  respectively.  However,  the separation of $f_1$ and $f_2$ is indicative of a deviation from the perfect circular symmetry. From the fitting curves of $f_1$ and $f_2$,  we can deduce $\lambda$ to be 3.56 and 4.13 for $f_1$ and $f_2$,  respectively.  Comparing it to the singular $\lambda$ of a circle, which is 3.832,  is indicative of an inherent source of circular asymmetry that could be from the aforementioned non-uniform drumhead deformation or eccentricity\cite{buchanan2005finite}.
\\\indent
The mechanical amplitude dependence on $V_{dc}$ when $V_{ac} << V_{dc}$ can be written as\cite{huang2018} (detailed derivation is shown in Supplementary Information):
\begin{equation}
z_{mn} = \frac{\sqrt{2\pi}\epsilon_0V_{ac}V_{dc}}{8\pi m_{eff}g_0^2\gamma_{mn}f_{mn}},
\label{eq:Equation_7}
\end{equation}
where $\epsilon_0$,  $m_{eff}$,  and $\gamma_{mn}$ are the values of the vacuum permittivity,  effective mass, and mechanical bandwidth, respectively.  As mentioned previously,  there is an asymmetry between the tuning of both the mechanical frequency and amplitude observed for the negative and positive voltages. This asymmetry can be attributed to the gate tunable nonlinearity arising from the electronic properties of the material, such as a multilayered graphite flake in this case\cite{Kim2018}. The gate affects the conductivity leading to the positive and negative asymmetry. This nonlinearity is also increased further by the drumhead non-uniformity as well\cite{tomi2015buckled}. This asymmetry is more apparent in the higher $(1,1)$ modes, where the increased number of the nodal lines inherently leads to smaller amplitudes, which then become more susceptible to this nonlinearity.
\subsection*{\label{sec:level2}Factors affecting tunability}
\indent
To further understand the mechanical frequency tunability, we vary the parameters of Equation 6 to compare the frequency tunability between graphite and another material, such as NbSe$_2$. Furthermore, we also vary the geometric parameters, such as thickness and radius and see the changes to the frequency tunability accordingly.  Figure 5a plots the relative frequency shift for both graphite and NbSe$_2$ resonators as a function of the device thickness while keeping all other parameters the same,  including the initial tension.  Figure 5b,  does the same as Figure 5a,  but the variation is in the device's radius.  From Figures 5a and 5b,  we see how these two geometrical parameters affect the frequency shift for both materials.  One can see that the frequency shift remains almost constant as the thickness increases until a certain value is reached,  then the frequency shift decreases.  On the contrary,  the frequency shift increases as the radius is increased,  but eventually plateaus when a certain value of radius is reached.  From these two dependencies,  which are both in agreement with Equation 6,  we can conclude that there is an optimal combination of the thickness and radius for the maximum frequency tunability. It is important to note here,  however,  that the differential motional capacitance is held constant for this comparison for simplicity.  In terms of material property difference between graphite and NbSe$_2$, the elastic modulus,  mass density and Poisson’s ratio affect both the stiffness and the mechanical frequency of the drum. If all geometric parameters are the same, then NbSe$_2$ should theoretically have better frequency tunability than graphite because it is less stiff just as shown in Figure 5. This means that the tensile stress induced by V$_{dc}$ would have less influence on the strain of the graphite drum as compared to the NbSe$_2$ drum. However, it is also important to note that because NbSe$_2$ is a denser material than graphite the corresponding mechanical frequency for NbSe$_2$ would be lower than that of graphite.  Overall,  the results of the calculations give important insights into how device parameters affect the frequency tunability (more detailed frequency tunability dependence on various parameters is shown in the Supplementary Information).  The next important step is to compare this device's frequency tunability with that of the reported devices with a similar geometry.
\begin{table}[htbp]
\centering
\begin{tabular*}{\textwidth}{ccccccccc}
 \hline 
 \hline 
 Device & Number of & $r$ ($\mu$m) & $f_0$ ($V_{dc}$ = 0)& $\Delta f$ (MHz)\footnotemark[1] & $g_0$ (nm) & $EF$ (V$\mu$m$^{-1}$)\footnotemark[2]& $f$ tunability & Optomechanical\\
   & Layers &   & (MHz) &   &   &   & per $EF$\footnotemark[3]  & Responsivity\footnotemark[4]\\
   &  &   &   &   &   &   &  (\%$\mu$m$V^{-1}$)  &  ($\mu$m$^{-1}$)\\
 \hline 
 Device A& 50 & 10 & 5.69 & 0.23 & 285 & 15.8 & 0.26 & 1.04\\
 Device  I\cite{barton2012photothermal} & 1 & 5.5 & 5.20 & 2.70 & 1370 & 7.30 & 7.11 & 0.27\\
 Device II\cite{singh2014} & 33 & 2 & 36.23 & -2.73 & 150 & 26.7 & -0.28 & 2.38\\
 Device III\cite{chen2013graphene} & 1 & 2 & 47 & 3.8 & 200 & 40 & 0.20 & 0.27\\
 Device IV\cite{weber2016force} & 1 & 1.6 & 67 & -27 & 85 & 35.9 & -1.12 & 0.27\\
 Device V\cite{DeAlba2016} & 1 & 4 & 8 & 4.5 & 1700 & 17.6 & 3.19 & 0.27\\
 Device VI\cite{mathew2016} & 1 & 1.75 & 30 & 20 & 300 & 125 & 0.53 & 0.27\\
  \hline 
  \hline 
\end{tabular*}
\caption{\label{tab:table1}Comparison of circular graphene nanomechanical resonators.  Important notes: 1) $\Delta f$ is $f$ at max $V_{dc}$ minus $f$ at $V_{dc}$ $\approx$ 0,  2) $EF$ is max $V_{dc}$/$g_0$,  3)  $f$ tunability is $\Delta f/f$($V_{dc}$ $\approx$ 0),  4) These values are calculated based on the device’s number of layers only from Figure 4.}
\end{table}
\\\indent
Table I shows the details of this comparison for 2D NMRs of a circular structure made from graphene.  For proper comparison,  the frequency tunability is normalized across devices by their respective electrical fields listed in the $EF$ column. The normalized values are shown in the frequency tunability column with the highest value belonging to Device I.  And while Device A does not give the highest value,  it is still not that far from most of the values for the other devices.  In order to appreciate the value of multilayered NMRs,  the predicted dependence of the optomechanical responsivity of these devices on the number of layers in Figure 4 is also calculated and is shown in the rightmost column.  Both frequency tunability and optomechanical responsivity are the figures of merit that need to be considered to determine the resonator's effectiveness as an optoelectrical transducer. Using these criteria,  both Device A and Device II could be considered as better optomechanical transducers than the other devices.  In other words,  these devices have the advantage of having a better optomechanical responsivity because of their multilayer nature without compromising their electrostatic tunability.  This,  in fact,  can be seen in Figure 5a,  where the device thickness can be increased up to a certain value before the frequency tunability falls off significantly.
\section*{Conclusion}
\indent
In summary,  we characterized nanomechanical resonators made from multilayered 2D materials and demonstrated their electrostatic tunability.  We showed that the optomechanical responsivity of the NMRs made from multilayers is greater than that of ultrathin ones. Furthermore, by using the multilayered NMR model,  we deduced the effect of various device parameters,  such as the thickness and radius of the drumhead,  on the mechanical frequency tunability. This model revealed that while the mechanical frequency tunability is highest for the thinnest resonators,  the resonator thickness can still be increased up to a certain value at which the tunability does not decrease much.  Together with the higher optomechanical responsivity,  this makes multilayered 2D NMRs the optimal choice for building optomechanical transducers.  These properties of multilayered 2D NMRs are important for the development of future coupled and integrated systems for various applications.
\section*{Author Contributions}
C.D.C. conceived the device and supervised the project; J.C.E. fabricated the samples; K.-H. L. and C.-Y. Y. designed and built the setup for optical measurements; J.C.E., M.A.C.A., and C.-Y.Y performed the measurements; J.C.E., M.A.C.A., J.Y.W., S.K., Y.P., and C.D.C analyzed the data, performed simulations, and wrote the manuscript; all authors discussed the results and contributed to the manuscript.
\section*{Conflict of Interest}
The authors declare no conflict of interest.

\section*{Acknowledgments}
We acknowledge the contributions of Tzu-Hui Hsu and Wen-Hao Chang to the fabrication of the devices and building the experimental setup.  We thank Bo-Ru Guo for her technical assistance.  We also thank the Taiwan International Graduate Program for the financial support. This project is funded by  Academia Sinica Grand Challenge Seed Program (AS-GC-109-08), Ministry of Science and Technology (MOST) of Taiwan (107-2112-M-001-001-MY3), Cost Share Programme (107-2911-I-001-511), the Royal Society International Exchanges Scheme (grant IES$\backslash$R3$\backslash$170029), and iMATE (2391-107-3001). We would also like to extend our gratitude for the Academia Sinica Nanocore facility.

\section*{Data Availability}
The data that support the findings of this study are available from the corresponding author upon reasonable request.

\newpage
\begin{figure*}
    \centering
    \includegraphics[width=1.0\linewidth]{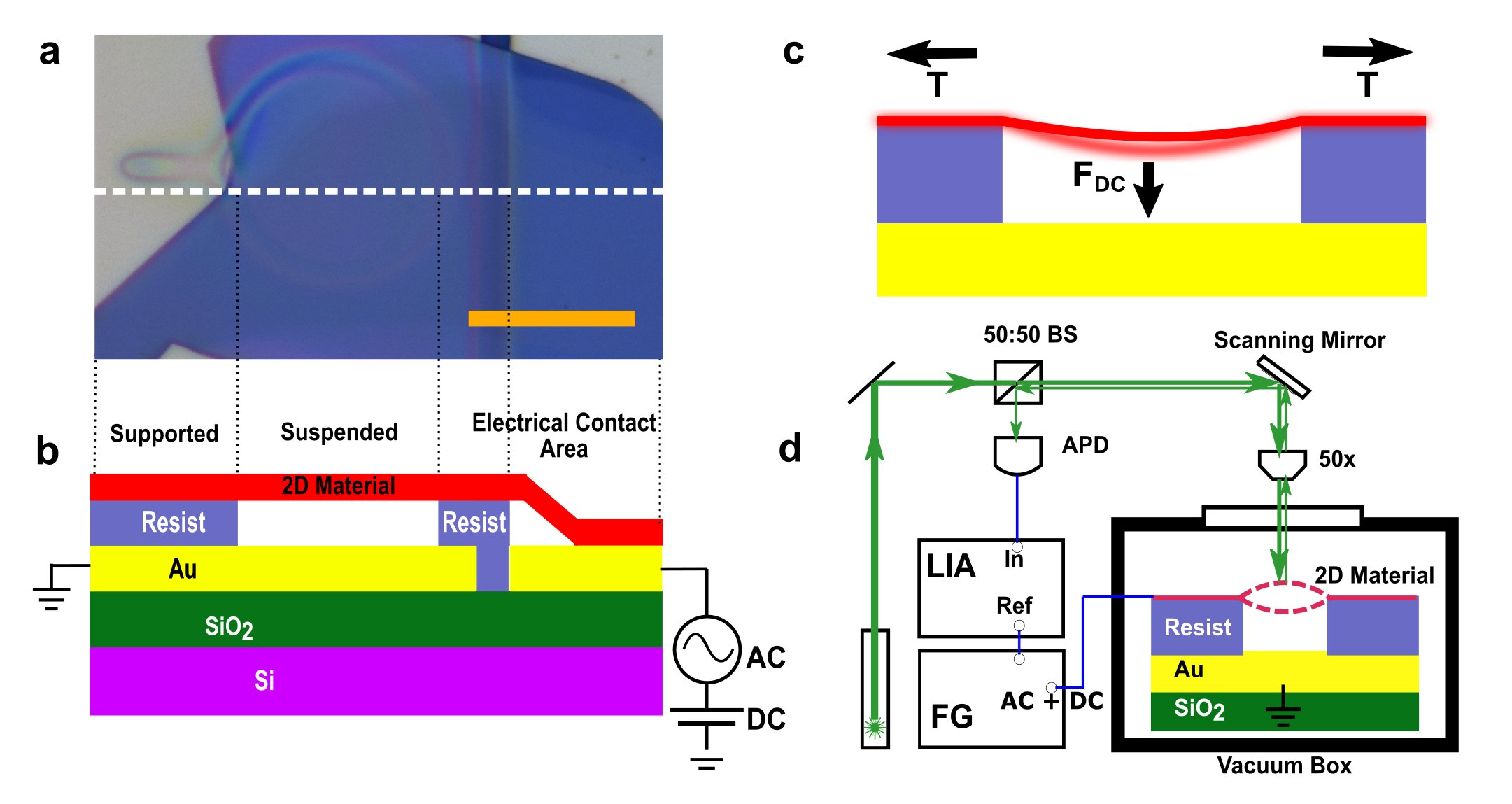}
    \caption{Device characterization.  (a) Optical image of the mechanical drum (scale bar is 7 $\mu$m).  (b) A schematic cross-section of the mechanical drum device along the dashed line in (a).  The graphite flake (red) is placed on top of a patterned 285 nm thick electron beam resist (blue), which serves as a spacer.  Two types of patterns are made.  The first one is the circular drumhead.  The second,  which is a few microns away from the circular drumhead,  is a larger rectangular opening,  where the flake is made to have electrical contacts with the gold bottom electrodes (yellow).  (c) A schematic of how the applied electrostatic force $F_{dc}$ is counterbalanced by the total radial tension $T$ in the drumhead boundary is shown.  (d) The sample is contained inside a vacuum box with a pressure of about $10^{-7}$ mbar.  The drum motion is detected using Fabry-Perot interferometry with a green laser (532 nm) and actuated using AC and DC voltages from the function generator. }
    \label{fig:Figure_1}
\end{figure*}
\begin{figure*}
    \centering
    \includegraphics[width=0.65\linewidth]{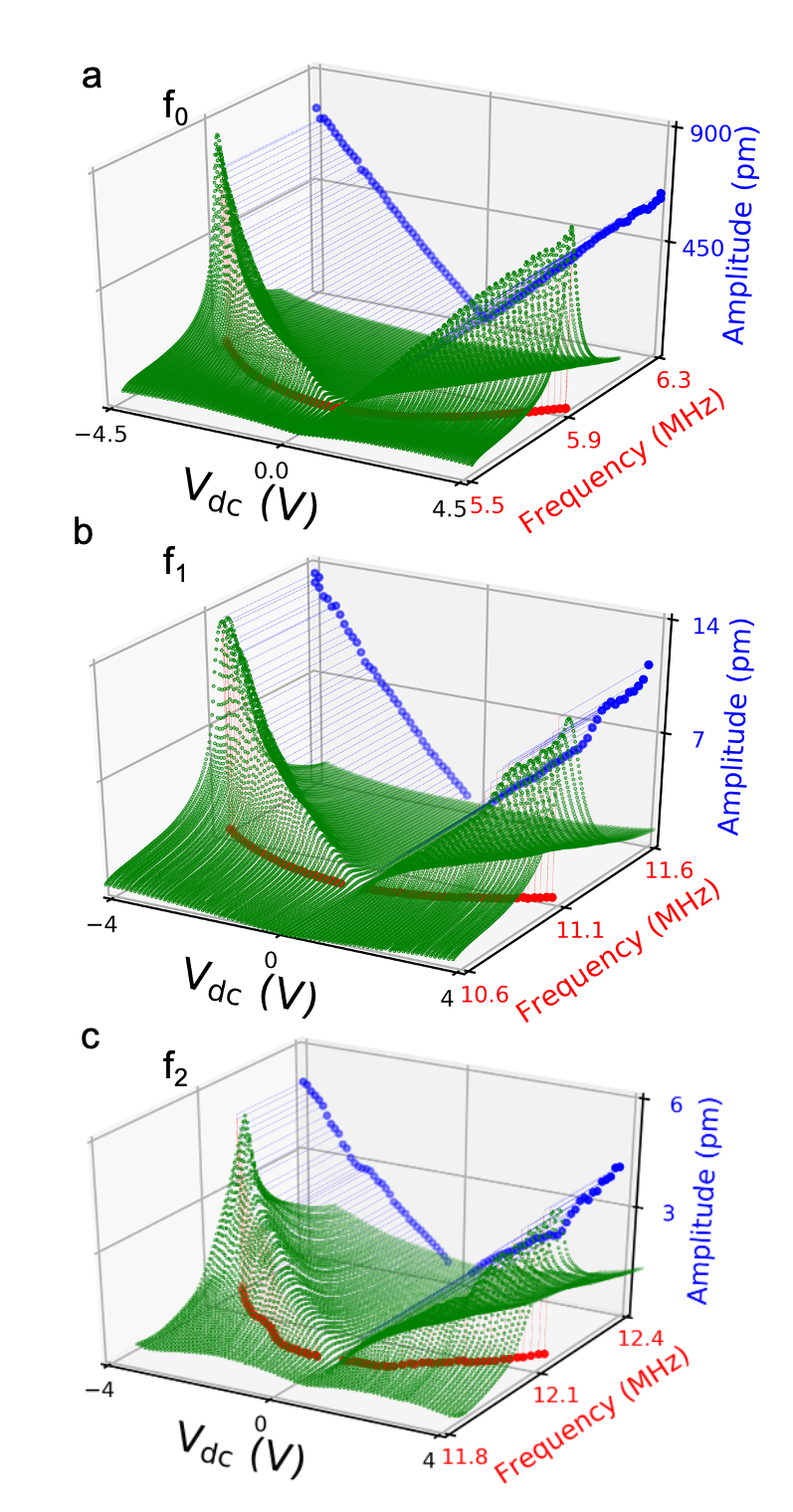}
    \caption{Electrostatic tuning.  Electrostatic tuning of the device (a) for the fundamental mode showing different driving frequency responses for different values of $V_{dc}$ from -4.5 V to 4.5 V at $V_{ac}$ = 2 mV$_{pp}$ and for the higher modes $f_1$ (b) and $f_2$ (c) showing different driving frequency responses for different values of $V_{dc}$ from -4 V to 4 V at $V_{ac}$ = 500 mV$_{pp}$.}
    \label{fig:Figure_2}
\end{figure*}
\begin{figure*}
    \centering
    \includegraphics[width=1.0\linewidth]{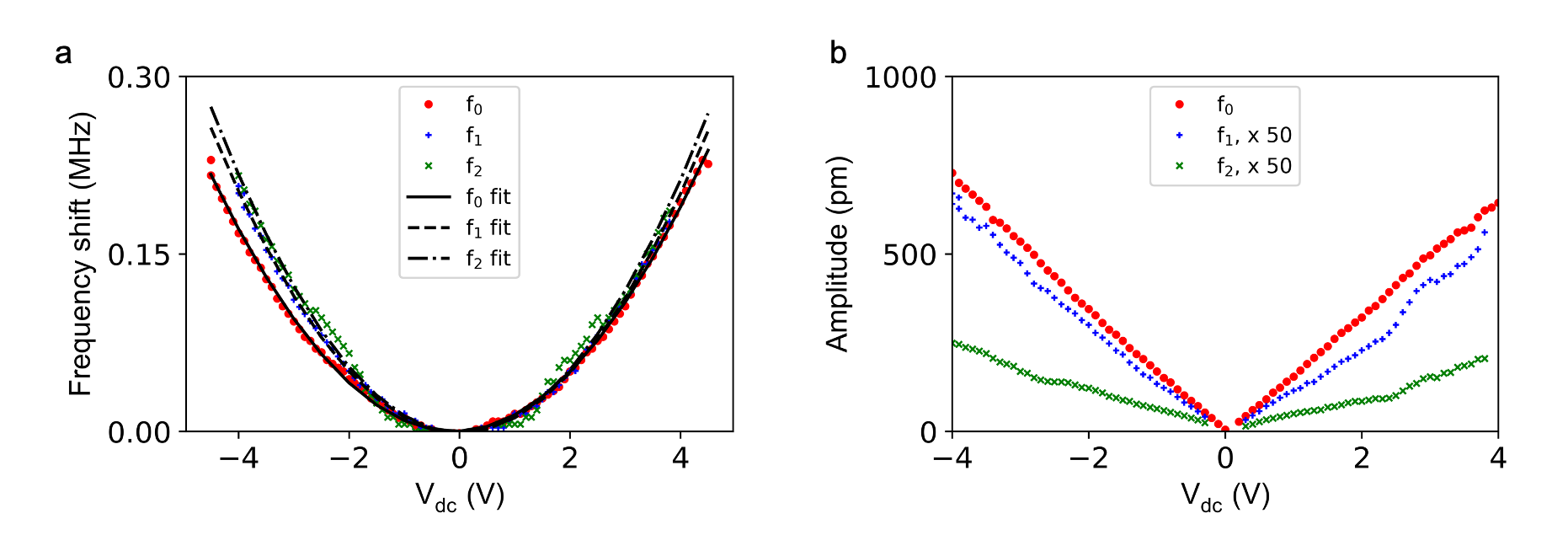}
    \caption{Comparison of electrostatic tunability of the mechanical modes of the graphite device.  (a) Frequency shift,  $\Delta f$,  of each mode as a function of $V_{dc}$ with the corresponding fits.  (b) Mechanical amplitude of each modes are shown as $V_{dc}$.  The amplitude of the higher frequency modes $f_1$ and $f_2$ are scaled by 50 for clarity.}
    \label{fig:Figure_3}
\end{figure*}
\begin{figure*}
    \centering
    \includegraphics[width=0.65\linewidth]{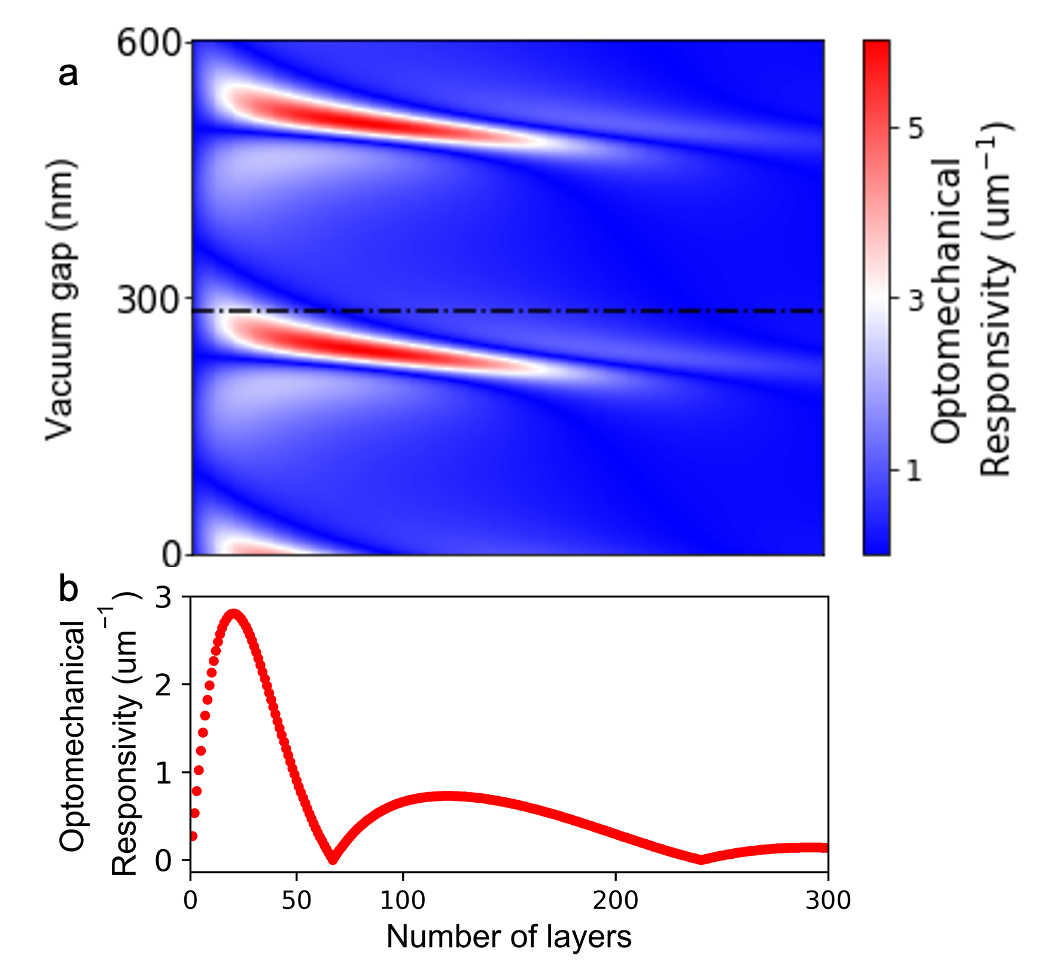}
    \caption{Calculated optomechanical responsivity. (a) Dependence of the optomechanical responsivity on the number of layers and spacer gap.  (b) Dependence of the optomechanical responsivity on the number of layers at a spacer gap of 285 nm,  as indicated by the black horizontal line in (a).  Conversion from the photodiode voltage response to $um^{-1}$ is done using the multiple interface calculations.}
    \label{fig:Figure_4}
\end{figure*}
\begin{figure*}
    \centering
    \includegraphics[width=1.0\linewidth]{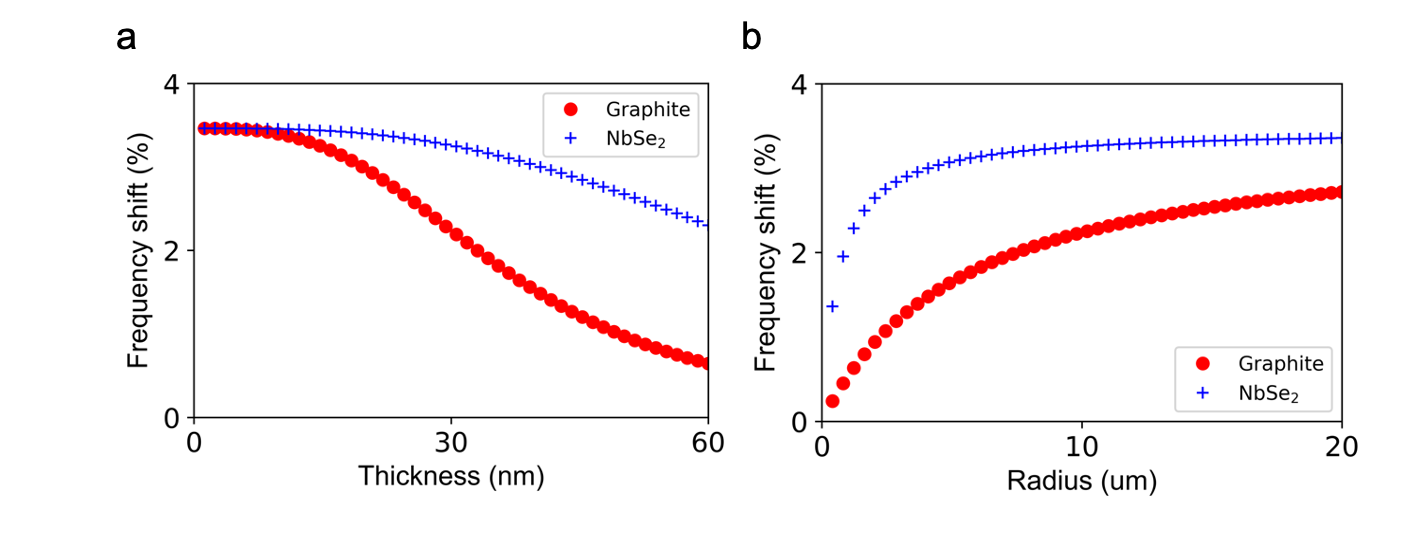}
    \caption{Comparison of electrostatic tunability using calculations from the multilayered NMR model. The electrostatic tunability is generated at max $V_{dc}$ = 4 V by varying the (a) thickness and (b) radius of the drumhead for both graphite and NbSe$_2$.  It should be noted that all other parameters are held constant including initial tension and differential motional capacitance,  while the radius for (a) is 10 um and the thickness for (b) is 30 nm.}
    \label{fig:Figure_5}
\end{figure*}
\end{document}